\documentclass[%
 reprint,
 amsmath,amssymb,
 aps,
 prl,
 floatfix,
 superscriptaddress,
]{revtex4-2}

\usepackage{graphicx}      % Required for including figures
\usepackage{dcolumn}       % Align table columns on decimal point
\usepackage{bm}            % Bold math
\usepackage{xcolor}        % Color support
\usepackage[normalem]{ulem}
\usepackage{mathrsfs}
\usepackage{algorithm}
\usepackage{algpseudocode}
\usepackage{chngcntr} 
\usepackage{hyperref}
\hypersetup{
    colorlinks=true,       % false: boxed links; true: colored links
    linkcolor=blue,        % color of internal links (change to black if preferred)
    citecolor=blue,        % color of links to bibliography
    filecolor=magenta,     % color of file links
    urlcolor=blue          % color of external links
}

\begin{document}

\title[Article Title]{Investigation of transverse instability in efficient plasma-based accelerators}

\author{Arohi Jain}
\email{arohi.jain@stonybrook.edu}
\affiliation {Department of Physics and Astronomy, Stony Brook University, Stony Brook, USA}
\author{Navid Vafaei-Najafabadi}
\email{navid.vafaei-najafabadi@stonybrook.edu}
\affiliation {Department of Physics and Astronomy, Stony Brook University, Stony Brook, USA}

\begin{abstract}
Simultaneously maximizing power-transfer efficiency and preserving beam quality in plasma-based accelerators is constrained by the transverse beam breakup instability. We present an analytical transverse wake model derived from the plasma wake potential that self-consistently captures deformed cavity boundaries under intense beam loading. This model maps efficiency limits against stability thresholds across a parametric scan to isolate an operating window. Three-dimensional particle-in-cell simulations validate the theoretical model, showing that the analytical centroid evolution matches numerical tracking during the acceleration of a tailored trapezoidal electron bunch over a 1~m propagation distance. The bunch achieves an energy gain of up to 16.5~GeV with nearly preserved transverse emittance, and a wake-to-trailing bunch power-transfer efficiency of nearly 80\%, with a final relative energy spread of less than 1.5\%. These results show that the efficiency-instability barrier can be circumvented through precise determination of transverse wake force and tailored beam loading, providing a design path toward compact, high-luminosity particle colliders.
\end{abstract}

\maketitle

Advanced accelerator concepts that employ plasmas as accelerating structures can sustain multi-GV/m gradients, orders of magnitude larger than those in conventional radio-frequency cavities. Consequently, plasma-based acceleration represents a highly promising pathway toward cost-effective, compact, high-energy particle colliders \cite{leemans2009laser, Schroeder2010, adli2022towards}. To be viable for such applications, however, plasma accelerator stages must produce high-luminosity beams at reasonable power consumption. Meeting this requirement demands high power-transfer efficiency while preserving the phase-space quality of the accelerated bunch, including low emittance and low energy spread \cite{benedetti2022whitepaper}.

% a plasma accelerator stage must simultaneously satisfy these requirements: it must operate at a high driver-to-trailing-bunch power-transfer efficiency, and it must preserve the transverse phase-space quality (low emittance and low energy spread) of the accelerated electron bunch to ensure high luminosity \cite{bane2003short, benedetti2022whitepaper}.

Recent theoretical work, however, has challenged the possibility of simultaneously fulfilling these requirements in a plasma accelerator in the highly nonlinear blowout regime \cite{lebedev2017efficiency}. In this regime, an intense laser pulse or a dense particle-beam driver completely expels the background plasma electrons, forming an ion cavity that provides focusing forces on the same order as the accelerating forces. These strong focusing forces can resonantly amplify small initial transverse offsets into large transverse oscillations, giving rise to the beam breakup (BBU) instability \cite{WhittumPRL,huang2007hosing}. An initial attempt to quantify this efficiency–instability trade-off was made by Lebedev \textit{et al.} \cite{lebedev2017efficiency}, whose analysis suggested that the instability becomes unavoidable at high power-transfer efficiency and that mitigation strategies such as BNS (Balakin–Novokhatsky–Smirnov) damping \cite{balakin1983vlepp} would compromise the beam quality needed for collider applications. In this paper, we reassess this proposed limitation by developing a theoretical description of transverse beam dynamics based directly on the underlying wakefield physics. We show that the earlier conclusion relied on simplifying assumptions that substantially overestimate the transverse force experienced by the trailing beam and, consequently, the severity of BBU-induced degradation. Using this framework, we map the parameter regime in which stable acceleration is maintained and show that low energy spread can be achieved simultaneously with stable beam transport and power-transfer efficiencies from wake to trailing beam on the order of 80\%.

The transverse dynamics of BBU instability can be formulated in terms of the centroid motion of individual bunch slices. For a slice located at longitudinal coordinate $\xi=z-ct$, the centroid $X(\xi,z)$, defined as the average transverse position of a bunch slice, evolves along the accelerator according to \cite{chao1993physics,SchroederPRL99,lehe2017,chen2020modeling}
\begin{equation}
\label{eq:motion}
\frac{\partial^2X}{\partial z^2}+\frac{1}{\gamma}\frac{\partial\gamma}{\partial z}\frac{\partial X}{\partial z}+ k_{\beta}^2X=-\frac{e}{\mathcal{E}} \tilde{F}_{\perp}, \quad (1)
\end{equation}
where  $k_{\beta}=k_{p}/2\gamma$ is the betatron wave number, $\mathcal{E} = \gamma(\xi,z) m_e c^2$ is the energy of an electron, $\gamma$ is the Lorentz factor, $k_{p}$ is the plasma wave vector. This equation describes the transverse motion of each slice under betatron focusing ($k_{\beta}^2X$), acceleration induced damping (second term) and transverse wake force $F_{\perp}$. Physically, each slice undergoes betatron oscillations in the ion-column focusing field with characteristic wavenumber ($k_\beta$). A transverse offset of the leading slices generates a wake force $\tilde{F}_\perp$ on the trailing slices and under uniform acceleration, the common betatron mode allows this intra-beam coupling to grow resonantly along the accelerator. 
% ; under uniform acceleration, the slices share the same natural betatron mode, allowing the intra-beam effect to grow resonantly over the acceleration length and amplify the slice-to-slice centroid displacement.

The transverse wake force per unit charge $\tilde{F}_{\perp}$ for a bunch slice with charge $q$ located at $\xi$ is given by:
\begin{equation}
\label{eq:f_perp}
\tilde{F}_{\perp}=\frac{F_{\perp}(\xi,z)}{q}= \int_{\xi}^{\xi_H} W_{\perp}(\xi'-\xi) \lambda(\xi') X(\xi',z) d\xi'. \quad
\end{equation}
Here, $W_{\perp}(\xi'-\xi)$ is the transverse wake function, $e$ is the electron charge, $\xi_H$ is the longitudinal position of the bunch head, $X(\xi',z)$ is the centroid of bunch slice located at $\xi'$, and $\lambda(\xi')=\int_0^{\infty} \rho(\xi',r) 2\pi r dr$ is the linear charge density of the bunch, with $\rho(\xi',r)$ representing the charge density of the accelerating electron bunch. Integration over $\xi'$ allows for the accumulation of the contributions of individual bunch slices to the total wake force experienced at $\xi$. 
\begin{figure}[t]
    \centering
    \includegraphics[width=1\linewidth]{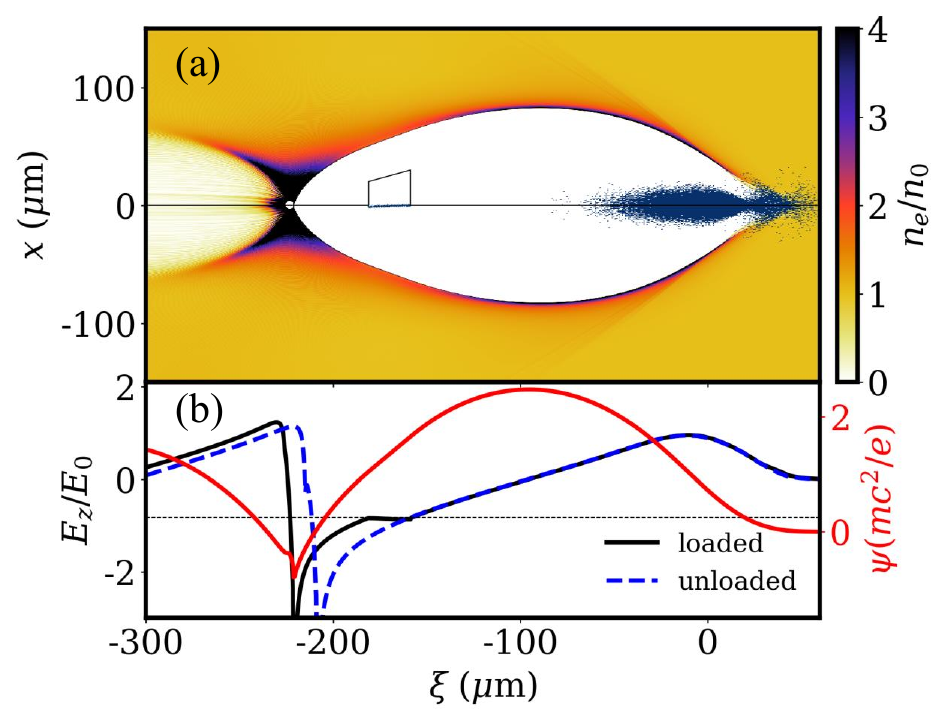}
    \caption{Simulation of an off-axis trailing bunch in an electron beam-driven plasma wake. (a) A 3.2 nC driver electron beam with  $\sigma_z=20$ $\mu$m  , and $\sigma_r=10$ $\mu$m  along with an off-axis electron bunch (offset $X_0=2$ $\mu$m) with charge $Q=1.1$ nC, $\sigma_z=23$ $\mu$m, and $\sigma_r=2$ $\mu$m propagating in a plasma with density $n_0=3\times10^{16}  \text{ cm}^{-3}$ ($E_0=17$ GV/m). (b) The longitudinal field of the trailing bunch partially cancels the unloaded field (dotted blue) of the bubble, producing a loaded field $E_t=0.82$$E_0$ (dotted black line) that is flattened in the region occupied by the bunch (black curve) along with wake function $\psi$ (red curve). }
    \label{fig:bubble wake with trapezoid beam}
\end{figure}

 %For low energy spread, we take a trapezoidal current profile (Fig. \ref{fig:bubble wake with trapezoid beam}). This distinctive shape, first proposed by M. Tzoufras \textit{et al}. in 2008 \cite{Tzoufras2008PRL}, balances the accelerating field across the entire bunch ensuring low energy spreads.

A precise determination of the transverse wake function, $W_{\perp}$, is crucial for accurate modeling of transverse instabilities in a plasma accelerator. The transverse wake function is linked to the longitudinal wakefield via the Short-Range Wake Theorem (Panofsky-Wenzel) \cite{fedotov1999transverse,bane2003short, lebedev2017efficiency}:
\begin{equation}\label{eq:wake_theorem}
    W_{\perp}  = \frac{2}{r_b^2  q} \int_{\xi}^{\xi'} E_z(u) \, du.
\end{equation}
Here, $E_z$ is the longitudinal electric field, $q$ is the bunch charge and $r_b(\xi')$ is the local bubble radius. Previous studies evaluated this field using non-linear wakefield theory \cite{lu2007generating}, assuming a rigid coupling between the bubble radius and the transverse force. 
Notably, Stupakov and Lebedev \cite{stupakov2018short,lebedev2017efficiency} modeled the transverse wake as
\[
W_\perp(\xi) \approx \frac{2\xi}{a(\xi)^4},
\qquad
a(\xi)=r_b(\xi)+\alpha k_p^{-1},
\]
where $a(\xi)$ was proposed as an effective ``iris’’ radius to account for the finite plasma skin depth and sheath thickness, and `$\alpha$' was treated as a numerical fit parameter, typically taken to be $\alpha \simeq 0.75$. This approach implicitly assumes that the bubble maintains a spherical or near-spherical topology even under load. While this assumption is reasonable for unloaded or lightly loaded wakes, in the high-efficiency regime, the high charge witness beam significantly deforms the bubble boundary, resulting in significant error in this model.

To address this, we moved beyond geometric approximations and derived the transverse force directly from the wake potential, defined as  $\psi(\xi)=\phi-A_z$ in normalized units, where $\phi$ is the scalar potential and $A_z$ is the $z$ component of the vector potential. Unless units are explicitly written or otherwise stated, in the remainder of paper, time and length are normalized to $\omega_p^{-1}$ and $k_p^{-1} =c/\omega_p$, respectively, velocity to the vacuum speed of light $c$, density to the electron density $n_0$, charge to the fundamental charge $e$, potentials to $m_ec^2/e$, linear charge density to $m_e c^2 \varepsilon_0/e$, magnetic field to $m_e\omega_p /e$,  and electric field to $m_ec\omega_p /e$. Here, $\omega_p=(n_0 e^2/m_e \varepsilon_0)^{1/2}$ is the plasma frequency and $m_e$ is the electron mass. Because of the direct dependence of the accelerating field on the wake potential through $E_z(\xi) = -\partial \psi / \partial \xi$, this formalism remains valid for arbitrary (including non-spherical) cavity geometries. The intra-beam effect of interest (i.e., instability of the witness bunch) arises exclusively from the wake
generated by the witness itself, whose contribution to the wake potential can be isolated \(\psi_{\text{witness}}(\xi) = \psi_{\text{total}}(\xi) - \psi_{\text{driver}}(\xi)\). Here, $\psi_{\text{total}}$ is the wake potential generated by both driver and trailing beams and $\psi_{\text{drive}}$ is the wake potential in the absence of the trailing beam. This leads to the wake potential model:
\begin{equation}\label{eq:wake-potentialmodel}
    W_{\perp}  = \frac{2}{r_b^2  q} [\psi_{\text{witness}}(\xi)-\psi_{\text{witness}}(\xi') ].
\end{equation}
This formulation corrects the previous
expression in two critical ways. First, Unlike the previous model, our model naturally incorporates the
exact, deformed boundary conditions of the loaded bubble without any additional fitting parameters. Second, The potential difference mathematically represents the exact integral of the longitudinal field experienced by the beam. The field of a beam loaded wakefield, for a beam occupying $\xi_H<\xi<\xi_F$, is given by
\begin{equation}\label{eq:Ez_full_loaded}
    % E_z(u) \;=\; E_t \;-\; \frac{\chi}{2}\,u,
    E_z^{\text{total}} =
    \begin{cases}
    -\dfrac{\chi \xi}{2}, & 0 < \xi < \xi_H, \\[6pt]
    E_t = -\dfrac{\chi\xi_H}{2}, & \xi_H < \xi < \xi_F,
    \end{cases}
\end{equation}
and the field for the unloaded wakefield is simply $E_z^{driver} =-\chi \xi/2$ over the wake. Here, $0<\chi\leq1$ accounts for the reduction of the longitudinal-field slope associated with the finite bubble radius $R_b<4$, which modifies the unloaded-wakefield slope from the ideal value $E_z^{\rm driver}=-\xi/2$.
\cite{yan2025study}. The corresponding potentials $\psi_{\text{total}}$ and $\psi_{\text{driver}}$ are then obtained by integrating the respective expressions for $E_z$. This approach yields an accurate transverse wake function $W_\perp$ in the beam-loaded regime, where the witness beam deforms the cavity and modifies the accelerating field, and provides the key input needed to predict BBU growth under high-efficiency acceleration conditions. Furthermore as the longitudinal dynamics and consequently $\psi$ are the result of beam current and are decoupled from transverse dynamics, $W_\perp$ does not change for a stable plasma bubble throughout propagation.

\begin{figure}[t]
    \centering
    \includegraphics[width=1\linewidth]{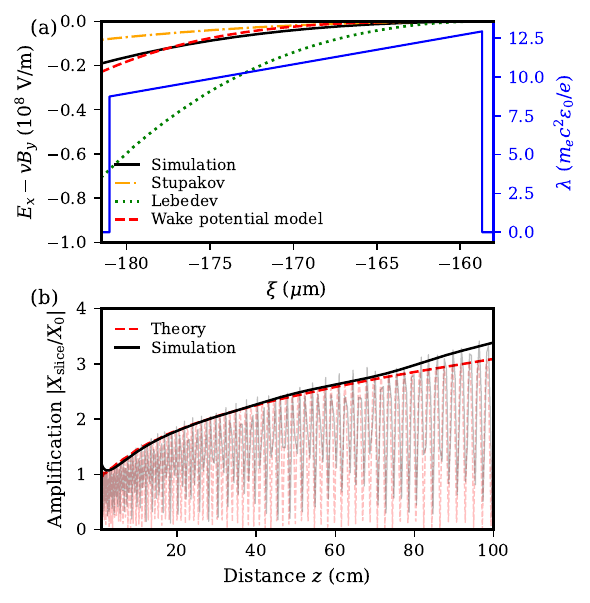}
    \caption{(a) The transverse force per unit charge, experienced by the trailing bunch. The analytical wake potential model (dashed red) shows excellent agreement with 3D HiPACE++ simulation results (solid black), outperforming previous models by Stupakov (dash-dotted orange) and Lebedev (dotted green). The bunch's linear charge density $\lambda$ is shown in a blue bold line. (b) Evolution of the BBU amplification factor at $\xi_{slice}= -180 \ \mu m$ over a 1~m plasma length. The bold red dashed (wake potential model) and solid black (simulation) curves represent the envelopes of the peak centroid oscillations, with the underlying light red and light black curves showing the centroid oscillations of the slice.}
    \label{fig:transverse force}
\end{figure}

%To investigate these dynamics in a high-efficiency regime that preserves pristine beam quality, we employ a specifically shaped trapezoidal trailing bunch designed to flatten the longitudinal accelerating field. Our theoretical model is rigorously validated against three-dimensional HiPACE++ particle-in-cell simulations, demonstrating remarkable agreement and significantly improved accuracy over previous transverse wake function models. 

We validate this formalism using three-dimensional HiPACE++ particle-in-cell (PIC) simulations \cite{diederichs2022hipace++} of a high-efficiency accelerating structure designed via beam-loading optimization \cite{yan2025study}. The parameters are selected to correspond to a blowout radius \(R_b\sim3\) in a regime that supports witness charges of order $1~\mathrm{nC}$ at a loaded accelerating field $>10$ GV/m, while providing a broad operating space in which high efficiency and transverse stability can be simultaneously explored. This operating point represents an accelerating module of high interest for future collider applications.
We use an electron drive beam based on realistic parameters (FACET facility \cite{hogan2010plasma} at SLAC National Accelerator Laboratory), for which the plasma bubble remains stable over meter-scale propagation, allowing the beam-breakup dynamics to be isolated from wake evolution. Although this choice avoids dephasing effects that can potentially suppress BBU growth, the formalism itself applies equally to laser-driven wakefield accelerators.
The simulation utilizes a moving window of $500 \times 500 \times 380\ \mu\text{m}$ with $2048 \times 2048 \times 800$ cells, propagating through a hydrogen plasma with upramp length of 1~cm and plateau density $n_0 = 3\times10^{16}\text{ cm}^{-3}$. 
The wake is excited by a $20\text{ GeV}$, $3.2\text{ nC}$ electron driver ($\sigma_x = \sigma_y = 10\ \mu\text{m}$, $\sigma_z = 20\ \mu\text{m}$) with normalized emittances of $\epsilon_{n,x} = 50\ \mu\text{m rad}$ and $\epsilon_{n,y} = 5\ \mu\text{m rad}$. 
%\textcolor{purple}{These drive beam parameters are closely matched to the capabilities of the FACET facility \cite{hogan2010plasma}, establishing a normalized blowout radius of $R_b \approx 2.7$ and a peak loaded field near $E_t = 1.0$. This configuration is chosen specifically to achieve an approximately $10\text{ GeV}$ single-stage energy gain, representing an accelerating module of high interest for future collider applications.} 
To minimize energy spread, a $1.1\text{ nC}$ trailing bunch with transverse dimensions $\sigma_x = \sigma_y = 2\ \mu\text{m}$ and emittances $\epsilon_{n,x} = \epsilon_{n,y} = 0.2\ \mu\text{m~rad}$ is modeled with a $23\ \mu\text{m}$ long tailored trapezoidal current profile \cite{Tzoufras2008PRL}. This distinctive shape balances the accelerating field across the entire bunch ensuring low energy spreads.
As shown in Fig.~\ref{fig:bubble wake with trapezoid beam}, an electron beam driver evacuates background plasma electrons to form a non-linear blowout cavity, within which the trailing bunch flattens the accelerating field $E_z$ across the trapezoidal profile.
To evaluate beam breakup (BBU) instability, the trailing bunch is simulated with an initial transverse offset comparable to its rms spot size, $X_0 =\sigma_r= 2\ \mu\text{m}$.

Using this simulation setup, we assess the predictive performance of our wake potential model against established theoretical frameworks. Figure \ref{fig:transverse force}(a) compares the transverse wake force predicted by different theoretical models against the PIC simulation result. The earlier model by Lebedev overpredicts the transverse wake by nearly a factor of four, while the Stupakov model underpredicts by a factor of two. Because the transverse wake force $\tilde{F}_{\perp}$ is obtained by integrating the wake function along the bunch, a factor-of-four error can lead to a substantial overestimate of the accumulated transverse instability. In contrast, our wake-potential model derived in Eq.~\eqref{eq:wake-potentialmodel} shows excellent agreement with simulation results. Figure \ref{fig:transverse force}(b) shows the resulting centroid amplification obtained from the PIC simulation compared with the one obtained by solving coupled differential equations describing the evolution of slice centroid and transverse force (Eqs. \ref{eq:motion} and \ref{eq:f_perp}), using the expression in Eq. \ref{eq:wake-potentialmodel} for the transverse wake function, $W_\perp$.
The agreement remains strong over a meter of propagation at an
accelerating gradient exceeding 10 GeV/m, demonstrating that the new formalism captures the transverse beam dynamics with predictive accuracy.

\begin{figure}[t]
    \centering
    \includegraphics[width=1\linewidth]{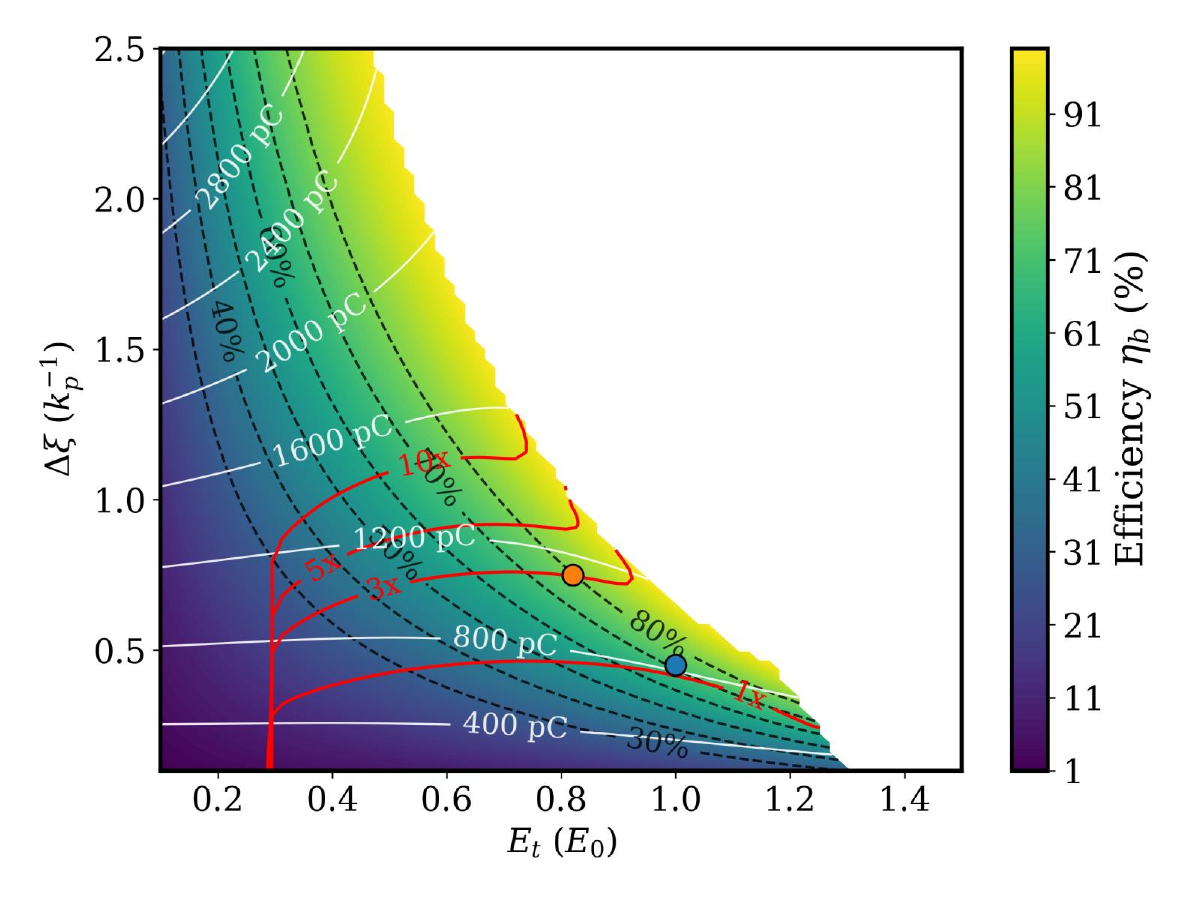}
    \caption{Parameter scan of the beam loading efficiency $\eta_b$ as a function of the loaded field $E_t$ and bunch length $\Delta \xi$, evaluated at a normalized bubble radius $R_b = 2.7$ and plasma density $n_0 = 3 \times 10^{16} \text{ cm}^{-3}$. Solid white curves indicate the bunch charge, and black dashed curves represent efficiency levels. The blank white region indicates the physically inaccessible parameter space dictated by the bubble closure constraint, ensuring that the cavity does not collapse before the end of the bunch. Red curves mark the stability thresholds defined by the maximum slice centroid ratio reaching limits of 1, 3, 5, and 10 for a final beam energy of 10 GeV. The blue and orange dots identify the $1\times$ and $3\times$ growth operating points, respectively, selected to study trailing bunch properties. The results shown in Fig. \ref{fig:bubble wake with trapezoid beam} and Fig. \ref{fig:transverse force} correspond to the  $3\times$-growth operating point indicated by the orange dot.}
    \label{fig:Parameter scan}
\end{figure}

Utilizing our analytical framework, we perform a comprehensive parametric study for a 10 GeV plasma stage with $R_b\sim3$, mapping efficiency limits against stability thresholds as functions of the loaded field, and the bunch length. Because our exact wake-potential formulation yields instantaneous solutions to the coupled centroid equations, it allows us to rapidly sweep across thousands of operational configurations to construct a predictive design map. To evaluate the trade-off between stability and accelerator performance, we next derive an expression for the energy extraction efficiency.

The charge per unit length for a bunch starting at $\xi_H$ needed to flatten the longitudinal field is given by\cite{Tzoufras2008PRL,yan2025study}:
\begin{equation}
    \lambda(\xi) = \lambda_0 - \frac{E_t}{\chi}(\xi - \xi_H), \quad \text{with} \quad \lambda_0 = \sqrt{\frac{R_b^4}{16} + \left(\frac{E_t}{\chi}\right)^4}.
\end{equation}
The overall efficiency of an accelerating stage can be expressed as the product of the driver-to-wake efficiency and the wake-to-trailing-bunch efficiency. Because the former depends on the specific driver and its evolution, we do not address it here. Instead, we focus on the wake-to-trailing-bunch energy transfer efficiency, which is determined by how effectively the trailing bunch beam-loads the wake and extracts the available wake energy. In what follows, $\eta_b$ refers to this wake-to-trailing-bunch energy-transfer efficiency, 
defined as the ratio of extracted energy to the maximum energy available in the wake,
given by $I_0=\pi \chi R_b^4/16$ \cite{yan2025study,Tzoufras2008PRL,Golovanov2021}:
\begin{equation}
    \eta_b = \frac{2\pi \int_{\xi_t}^{\xi_f} E_t \lambda(\xi) d\xi}{\chi\pi R_b^4 / 16}.
\end{equation}
Integrating the charge distribution $\lambda(\xi)$ from the bunch head ($\xi_H$) to the bunch tail ($\xi_f$) yields:

\begin{equation}
\label{eq:efficiency_final}
    \eta_b = \frac{32 E_t \Delta\xi}{\chi R_b^4} \left[ \sqrt{\frac{R_b^4}{16} + \left(\frac{E_t}{\chi}\right)^4} - \frac{E_t \Delta\xi}{2\chi} \right].
\end{equation}

\begin{figure}[t]
    \centering
    \includegraphics[width=1\linewidth]{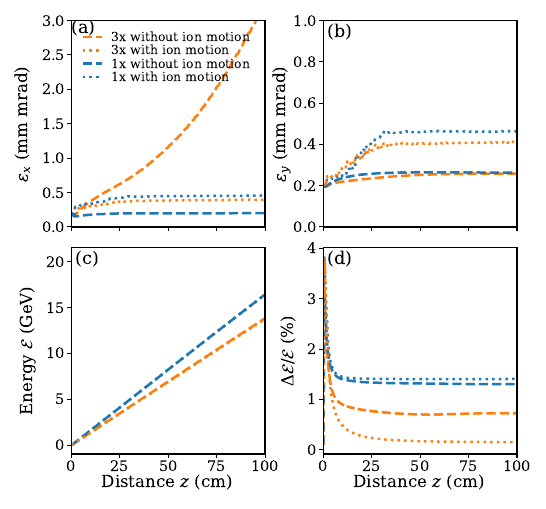}
    \caption{ Evolution of trailing bunch parameters over a 1~m propagation distance, demonstrating high beam quality at the operating points identified by our analytical parameter scan. The optimized 1x growth design (blue, corresponding to blue dot from Fig. \ref{fig:Parameter scan}) and 3x growth design (orange, corresponding to orange dot from Fig. \ref{fig:Parameter scan}) are modeled with ion motion (dotted) and without ion motion (dashed). (a), (b) Normalized transverse emittances $\epsilon_{x}$ and $\epsilon_{y}$. (c) Mean longitudinal energy gain, where the higher $E_t$ of the $1\times$ design yields a maximum gain of 16.5~GeV. (d) Relative energy spread $\Delta\mathcal{E}/\mathcal{E}$ maintained tightly below 1.5\% across all configurations.}
    \label{fig:beam_evolution}
\end{figure}

This closed-form model enables simultaneous optimization of energy extraction and BBU stability as a function of the target loaded gradient $E_t$, maximum bubble radius $R_b$, and bunch length $\Delta \xi$. 
The parameter scan in Fig.~\ref{fig:Parameter scan} identifies high-efficiency operating points where the predicted instability growth remains bounded. 
By evaluating the maximum slice centroid ratio across the bunch, the model defines thresholds for tolerable beam-slice amplitude amplification. 
This allows us to select an operational point that maximizes both beam charge and efficiency while limiting transverse oscillations growth. Specifically, to demonstrate operation within this high-efficiency, low-instability regime, we highlight two optimal configurations for the study of bunch properties evolution (as shown in Fig.~\ref{fig:beam_evolution}): a 1$\times$ growth design (blue dot; $E_t = 1.0$, $\Delta\xi = 0.45$, 0.8~nC, $\eta_b=72 \% $ ) that maximizes the accelerating gradient while maintaining strict baseline stability, and a 3$\times$ growth design (orange dot; $E_t = 0.82$, $\Delta\xi = 0.75$, 1.1~nC, $\eta_b=80 \%$) that extracts higher beam charge at greater overall efficiency while still reasonably bounding instability growth. The results presented in Fig. \ref{fig:bubble wake with trapezoid beam} and \ref{fig:transverse force} correspond to the latter case.

%Without ion motion, the 1x design preserves the initial 200~nm~rad emittance. With ion motion included, phase-mix detuning successfully clamps the otherwise unstable 3x design, establishing a stable operating floor below 0.5~mm~mrad for both configurations.

Finally, we evaluate the preservation of the accelerated beam quality for the two selected configurations over a meter-long propagation distance. The 1$\times$ design exhibits pristine stability, preserving the initial normalized transverse emittance of (0.2~mm~mrad) over the \(1~\mathrm{m}\) plasma stage and confirming the effectiveness of the stability-constrained design. In comparison, the \(3\times\) design allows modest centroid amplification in exchange for higher charge and greater wake-to-trailing-bunch efficiency, which results in a gradual emittance growth.
%%%%%%%%%%%%%%%%%%%%%%%%%%%%%%
Even this gradual growth, however, is mitigated once ion-motion physics is included. The ion response provides a well-known mechanism for bounding BBU growth in plasma accelerators \cite{Mehrling-ion2018} and modifies the transverse dynamics in two ways. First, it produces a nonuniform focusing force along the trailing bunch, causing different slices to accumulate different betatron phases. This phase mixing detunes the otherwise coherent instability, reducing the BBU growth rate. Second, the ion response introduces additional transverse forces beyond those shown in Fig. \ref{fig:transverse force}, which can modify the projected beam quality even in configurations with limited centroid growth. As shown in Fig. \ref{fig:beam_evolution}, the net effect is that ion motion bounds the emittance growth of the \(3\times\) design, while also introducing a baseline emittance increase that brings both selected configurations to normalized transverse emittances just below \(0.5~\mathrm{mm\ mrad}\).
 % the ion-induced phase mixing bounds the instability and reduces the associated emittance growth. At the same time, ion motion introduces a baseline emittance increase, bringing both selected configurations to normalized transverse emittances below \(0.5~\mathrm{mm,mrad}\). 
 Moreover, because the strength of ion motion depends primarily on the trailing-bunch density, and therefore on its transverse size, it represents an additional optimization axis that is independent of the beam-loading and efficiency physics considered above.

% Figure~\ref{fig:beam_evolution} illustrates the evolution of the trailing bunch parameters extracted from the 3D PIC simulations. As shown in Figs.~\ref{fig:beam_evolution}(a) and (b),the modest increase in the oscillation amplitude for the  3x case results in steady (small amplitude) growth in emittance. However, because the amplification is small, this effect is easily overcome by ion motion, and the accelerator is able to maintain the normalized transverse emittances ($\epsilon_{x}, \epsilon_{y}$) below 1~mm~mrad across the 1~m plasma accelerator. 

% \textcolor{purple}{Conversely, in the absence of ion motion, the 1x design exhibits pristine stability, completely preserving the initial 0.2~mm~mrad emittance. When realistic ion motion is introduced, the non-linear focusing forces of the ion column induce phase-mix detuning; while this mechanism bounds the growth of the $3\times$ case, it simultaneously introduces a baseline emittance growth that brings both configurations below 0.5~mm~mrad.
Concurrently, Figs.~\ref{fig:beam_evolution}(c) and (d) track the longitudinal bunch properties. Driven by the flattened accelerating gradients, the mean longitudinal energy increases linearly to 14~GeV for the 3$\times$ design and up to 16.5~GeV for the 1$\times$ design [Fig.~\ref{fig:beam_evolution}(c)], while the relative energy spread remains tightly bounded below 1.5\% across all modeled configurations [Fig.~\ref{fig:beam_evolution}(d)]. Crucially, excellent beam quality is preserved together %these pristine beam qualities coexist 
with exceptional energy transfer metrics: the high localized beam-loading efficiencies ($\eta_b = 73\%$ for 1$\times$ and $83\%$ for 3$\times$). This confirms that near-maximum wake energy extraction can be simultaneously sustained alongside excellent emittance preservation and narrow energy spreads.

 For completeness, we also evaluate the overall drive-to-trailing-bunch power-transfer efficiency in the simulations, obtaining \(\eta=34.2\%\) and \(39.4\%\) for the 1$\times$ and 3$\times$ design cases, respectively. The remaining energy fraction is accounted for by the initial driver-to-wake excitation coupling efficiency, which was not optimized in the present simulations and therefore represents an additional axis for future efficiency gains, for example through tailored driver profiles such as ramped-density beams \cite{svensson2026}.

% The remaining energy fraction is accounted for by the initial driver-to-wake excitation coupling efficiency and the residual, un-extracted wakefields left behind in the plasma column following the passage of the trailing bunch. 

In conclusion, three-dimensional particle-in-cell simulations demonstrate excellent agreement with our analytical predictions of centroid oscillations and transverse wake forces. The reults of the simulations confirm the validity of this model, verifying a stable operational window where a near $80\%$ energy transfer efficiency is achieved simultaneously with sub-0.5 mm mrad emittance preservation, an energy gain of up to 16.5~GeV, and a $\textless 1.5\%$ relative energy spread over a 1~m acceleration distance. By moving past rigid geometric approximations to establish an exact wake-potential formulation, this work provides a framework to map out optimized accelerator configurations that maximize power transfer without compromising beam stability. Consequently, these results offer a clearer, actionable design path that strengthens the foundational physics required for future compact, high-luminosity particle colliders. Finally, the wake-function framework introduced here provides a natural path for incorporating additional plasma effects, including ion motion, in future work.

\section*{Acknowledgments}
We acknowledge the support by the U.S. Department of Energy, Office of Science under Award No. DE-SC-0014043, DE-SC-0020396, DE-SC-0024277, NSF CAREER Award PHY-2238840,  and resources of NERSC facility operated under contract No. DE-AC02-5CH11231.

\bibliography{references}

\end{document}